# 이벤트 재구성을 위한 타임스탬프 갱신 임계치

# Update Thresholds of More Accurate Time Stamp for Event Reconstruction

조슈아 제임스[*], 장윤식[**]

**Joshua I. James[*], Yunsik Jang[**]**

**요 약** 용의자가 어떤 행위를 했는지 특정하기 위한 경우와 같이 디지털 조사에서 특정한 행위나 이벤트의 발생시간을 확인하기 위해 타임스탬프에 의존하는 시스템이 많다. 하지만 객체의 갱신은 실제 이벤트의 발생시점보다 약간의 시간차를 두고 이루어지게 된다. 이 논문에서는 타임스탬프와 관련된 객체를 가진 디지털시스템의 간단한 모델을 정의한다. 이 모델은 타임스탬프와 관련된 객체의 갱신 패턴을 예측하는데 사용되며 갱신 시간차 범위에 대한 예측을 가능하게 한다. 경험적 연구를 통해 타임스탬프 갱신패턴이 동시적이지 않다는 것을 보이고 특정한 시스템에서 보다 정확한 행위시점을 결정하기 위한 타임스탬프 갱신 분포를 계산하는 방법을 제시한다.

**Abstract** Many systems rely on reliable timestamps to determine the time of a particular action or event. This is especially true in digital investigations where investigators are attempting to determine when a suspect actually committed an action. The challenge, however, is that objects are not updated at the exact moment that an event occurs, but within some time-span after the actual event. In this work we define a simple model of digital systems with objects that have associated timestamps. The model is used to predict object update patterns for objects with associated timestamps, and make predictions about these update time-spans. Through empirical studies of digital systems, we show that timestamp update patterns are not instantaneous. We then provide a method for calculating the distribution of timestamp updates on a particular system to determine more accurate action instance times.

## Ⅰ. Introduction

Many systems rely on reliable timestamps to determine the time of a particular action or event. This is especially true in digital investigations where investigators are attempting to determine when a suspect actually committed an action. The challenge, however, is that objects are not updated at the exact moment that an event occurs, but within some time-span after the actual event. This time-span differs from system to system, depending on processing power, current load, type of process, etc. The result is that the object time stamp update process is not instantaneous. This sometimes make is difficult to differentiate two different actions that happened in a system – or their order – due to delays in time stamp updates. In order to accurately differentiate between multiple action instances, trace update duration must be defined for the particular action. In this work we define a simple model of digital systems with objects that have associated timestamps. The model is used to predict object update patterns for objects with associated timestamps, and make predictions about these update time-spans. Through empirical studies of digital systems, we show that time stamp update patterns are not instantaneous. We then provide a method for calculating the distribution of time stamp updates on a particular system to determine more accurate action instance times.

[*]정희원, 한림대학교 국제학부
[**]정희원, 한림대학교 국제학부 (교신저자)
접수일자 2017년 2월 23일, 수정완료 2017년 3월 23일
게재확정일자 2017년 4월 7일

Since time stamps are so important for information security and digital forensic investigations, many prior works have looked at how time stamps can be used to reconstruct events. Willassen[1][2] focuses on hypothesis-based approaches for time stamp correlation. Essentially analysis and verification of time stamps. Similarly, Carrier[3] used hypothesis based methods, but focused more on the state of the system to define events. Gladyshev and Patel[4] also used state machine analysis for event reconstruction in systems, allowing investigators to generate all possible paths through a state machine (digital system) and compare hypotheses with the available state space.

Other methods focused less on the state space of the system, and more on the direct observation of time stamps. For example, Koen and Oliver [5] give an overall account of how time stamps can be used for digital investigation purposes that is still commonly used in many digital systems[6]. James and Gladyshev[7][8] go further by giving a method for the detection of action instances based on time stamp update patterns. The challenge with time stamp analysis in postmortem investigations is the potential lack of data, especially as the investigator attempts to reconstruct older events. While there are several methods for filling in the gaps of time information[9], there is still more work to be done on aligning time stamps from different systems, and analyzing complex events with incomplete time information[10][11].

While all of these prior methods attempt to determine actions in time, none have identified the challenge of varying time stamp update times over a given threshold. The result is that investigators tend to take the observed time stamp value as the exact time that an action occurred, which may not be correct.

## 1. Contribution

This work contributes to the area of time stamp analysis for information security and digital forensic investigations by demonstrating that time stamp updates during an action instance is not instantaneous. This has implications for the way investigators think about when actions took place, with the actually occurrence of the action potentially happening minutes before the recorded time.

## II. TRACE-BASED DIGITAL INVESTIGATIONS

Prior works have explored the use of time stamps in digital forensic investigations. Time stamps are often used for temporal ordering in event reconstruction. A time stamp update in a system can be observed by an investigator. This trace can be associated with a number of hypotheses to determine whether it supports or denies a claim[?][?]. A trace is defined here as any object that can be observed that supports or denies some hypothesis[?].

## 1. Action Instances

In a computer system, actions cause processes to execute over an unknown period of time that modify or create objects as they are being executed. Multiple actions can start multiple concurrent processes, and multiple processes can modify the same objects. We can define actions as having an almost immediate effect on the system, where all effects on the system happen in a short period of time from each other. This means that when an action happens, related object meta-data is updated within a short period of time after the action actually occurs.

An action instance is defined as any occurrence that updates a collection of traces on a given system. Two actions may be functionally equivalent if both action instances produce the same pattern of trace updates.

## 2. Detection Action Instances

Similar to[12], action instances can be detected by using meta-data grouping over a set period of time. This trace update threshold is variable. For example, a collection of traces may be updated quickly when no other processes are running on the system, and slowly when processor load is high. Further, the update threshold is hardware specific. The speed of the processor, disk, bus, etc. will affect the trace update threshold. For example, a user visits a website and downloads the web page to cache. Traces are updated on disk over several tens of seconds, and the user reads the content. In this case, traces are updated on disk within 1 to 2 minutes, and are all related to the general action "browse the Internet". After trace updates, there are no updates for another few seconds. We could select this gap as an action instance cut-off, but then if the user is actively searching for multiple things, no gap would be found. The goal of this work is to determine a correct, generally applicable time stamp update threshold.

## III. OBJECT TIME STAMP UPDATE THRESHOLD

The object time stamp update process is not instantaneous. In order to accurately differentiate between multiple action instances, trace update duration must be defined for the particular action. The trace update times, in seconds, of the action instances "Open Internet Explorer 8" and "Open Firefox 3.6" were surveyed on 25 computer systems running Windows XP or Windows 7, with results shown in Figures 1 and 2. The results show that action instances' update duration will be different depending on the hardware of the system, as well as the state of the software. Because the time for a trace to be updated is variable, it must be described as a range. From experimentation, it was determined that the object update times may be modeled as a normal distribution. A standard deviation ($\sigma$) of $2\sigma$ was chosen as the standard threshold limiter to attempt to reduce unlikely outliers. This decision was made based on the fact that if the threshold is too large, then multiple instances of an action may be considered as one instance. A $2\sigma$ limit will cover approximately 95% of the distribution, effectively allowing outliers to be detected as multiple instances of the same action.

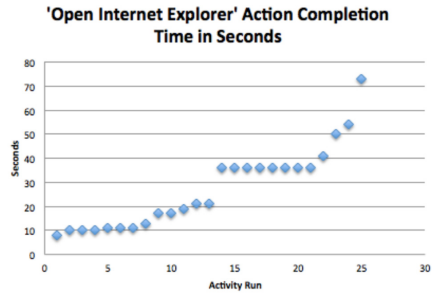

Fig. 1. Graph of the time in seconds it took for the action 'Open Internet Explorer' to complete on the tested system ordered from shortest to longest run

그림 1. 가장 빠른 실행순으로 테스트 시스템에서 Internet Explorer 실행에 걸린 시간(초)

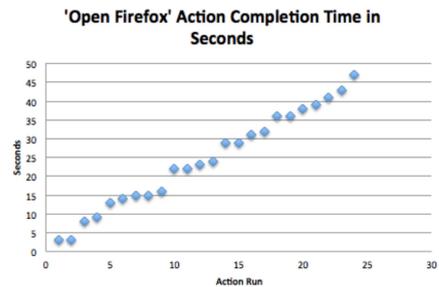

Fig. 2. Graph of the time in seconds it took for the action 'Open Firefox' to complete on the tested system ordered from shortest to longest run

그림 2. 가장 빠른 실행순으로 테스트 시스템에서 Firefox 실행에 걸린 시간(초)

For the action "Open Internet Explorer 8", the average trace update duration was 27.4 seconds, with a standard deviation of 16.76 seconds. The update threshold with a 2σ limiter is from 0 to 61 seconds. Figure 3 shows a histogram of then given data specifically for the action 'Opening Internet Explorer'. From Figure 3 it can be seen that update durations become fewer as time increases. In this case, the majority of update durations took place between 8 and 42 seconds after the action instance. After which there was a decline in the number of update durations per interval, with no update duration that lasted longer than 76 seconds.

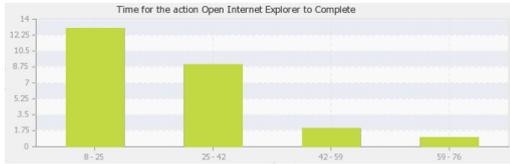

Fig. 3. Histogram of Internet Explorer update interval times in seconds where the X axis is time in seconds and the Y axis is the number of occurrences within the update duration
그림 3. X축은 시간(초)이고 Y축은 업데이트 기간 내 발생 횟수를 보여주는 Internet Explorer 업데이트 간격 시간

By modeling the data as a normal distribution, a standard threshold limiter (Θ) can be calculated, which, in the case of Opening Firefox, limits the maximum update threshold to 50 seconds.

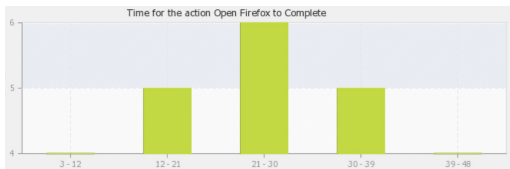

Fig. 4. Histogram of Firefox update interval times in seconds where the X axis is time in seconds and the Y axis is the number of occurrences within the update duration
그림 4. X 축이 시간 (초)이고 Y 축이 업데이트 기간 내의 발생 횟수인 Firefox 업데이트 간격 시간(초)

## 1. Action Instance Time Span Approximation

With knowledge of the object update threshold associated with a particular action, the time of the action instance may be approximated based on the associated object time stamp values. Objects are associated with action instances through observation[7]. Once objects are associated with a particular action instance, and have been categorized by their update patterns, the time span in which the action instance must have happened can be approximated.

First, each time stamp value in the set of returned time stamps is sorted from oldest to newest. For all objects where difference in time starting from the oldest to newest returned time stamp value is less than or equal to the action instance update threshold, these objects are grouped. The approximate time span of the action instance that updated each object is greater than or equal to the most recently updated (newest) time stamp in the set of grouped objects minus the action instance update threshold, and is less than or equal to the least recently (oldest) time stamp in the set of grouped objects.

For example, an action instance associated with time stamps t1 and t2 may be approximated based on the maximum action instance threshold where the instance must have occurred in the timespan before the least recent time stamp t1 and most recent time stamp t2 minus Θ. The time-span in which the action instance may be bound is denoted as $(t2 - Θ) \leq i.\tau \leq t1$ where $i.\tau$ is the time-range of the action instance. This time-bounding method to approximate the time of the action instance is shown in Figure 5.

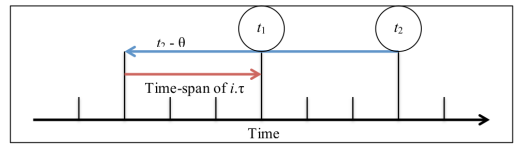

Fig. 5. Action instance time-span approximation based on time bounding before the least recently updated time stamp (t1) and the most recently updated time stamp (t2) minus the action's associated update threshold ($\theta$).
그림 5. 가장 최근 업데이트 된 타임스탬프(t2)와 가장 오래된 엡데이트 타임스탬프(t1)에서 행위의 관련 업데이트 임계값($\theta$)을 뺀 시간 경계에 기반한 행위 인스턴스 시간 범위 근사치

Using this method, approximation of the time-span in which each action instance must have occurred may be determined. However, if the trace update time lies within the object update threshold of multiple actions, then determination of which specific instance updated the time stamp is impossible.

## IV. EXPERIMENTATION

To determine trace update behavior for classification purposes, time stamps for objects related to pre-set actions were collected. Each action was executed 10 times. Other non-related actions were executed at least 2 minutes after the action of interest, including system shutdown and startup actions to introduce noise. After each execution of the action and noise-producing session, time stamps of all previously identified action-associated objects were collected as Fig. 6 and 7.

For each tested action, the results of each instance of the action were analyzed to determine trace category association[13]. This process consisted of comparing the trace time with the known execution time. For testing, an initial object update threshold of 120 seconds was assigned based on the observation that past action instance update thresholds were normally within 60 seconds. A threshold of 120 seconds allows for initial exploratory analysis that will be made more specific in later examination. Using this initial threshold, action instance traces were categorized according to their category. Description of categories of time stamp updates have been described in[13], and are out of the scope of this paper.

Once categorization was complete, a second level of refinement was required to calculate a more accurate object update threshold, and verify traces were correctly categorized. The process was run another 10 times, each time examining the update times compared with the known execution time. For both actions, the object update threshold was lower than the initial 120 seconds threshold. However, traces were not usually re-categorized because of the lower threshold – the time between action instances was sufficient to differentiate between executions, even with a longer threshold. Using the derived associated trace list, the object update threshold derivation process described in previously must be sampled on many machines to attempt to get a representative update threshold. The update threshold for IE8 was found to be 61 seconds, and the update threshold for FF3 was found to be 50 seconds.

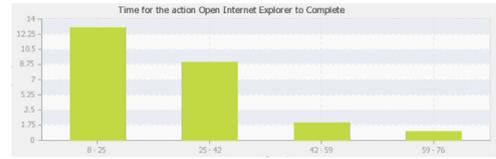

Fig. 6. Histogram of Internet Explorer update interval times in seconds where the X axis is time in seconds and the Y axis is the number of occurrences within the update duration
그림 6. X 축의 시간 (초)이고 Y 축은 업데이트 기간 내의 발생 횟수인 Internet Explorer 업데이트 간격 시간

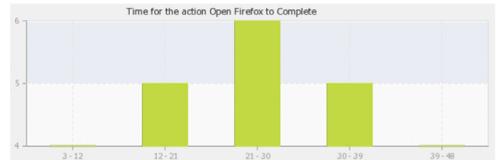

Fig. 7. Histogram of Firefox update interval times in seconds where the X axis is time in seconds and the Y axis is the number of occurrences within the update duration
그림 7. X 축이 시간 (초)이고 Y 축이 업데이트 기간 내의 발생 횟수인 Firefox 업데이트 간격 시간

This threshold can now be considered representative for the range of time that updates are written to disk after the action takes place. From here update thresholds could be created per application, or a general range could be created for a specific type of system (Windows, Linux, etc.).

### 1. Discussion

The proposed method shows that the actual time that an action is initiated, and the time that observable traces are updated has a delay. Investigators should not take time stamps as the exact time that an event occurred, but rather an approximation of the time an event took place. In our experiments we found that traces were created in a range of about 1 minute, but some traces were created much longer. This means a suspect could potentially start a process, and immediately leave the computer. Traces may be created after the suspect left the room, at least leaving reasonable doubt whether the suspect actually committed the crime. Using the proposed method, we can explain seeming discrepancies in time stamp updates on a system, and potential real-world user actions.

## V. CONCLUSIONS

This work showed that time stamp updates associated with action instances are not instantaneous. Investigators should be aware that there is a range of time in which updates are saved to disk. This range can be calculated experimentally for a specific system, or a general range can be calculated for a group of similar actions on similar systems. For example, opening a browser in a Windows computer. Investigators can use this time range to help explain potential defensive discrepancies that a suspect may claim in court. By calculating a general time range for specific actions, investigators can more accurately determine whether the suspect could have executed the action instance.

## 저자 소개

**Joshua Issac James(정회원)**

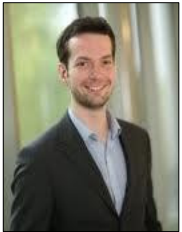

- 2013년 9월 : University College Dublin, Dublin, Ireland. PhD Computer Science by research in Digital Forensic Investigation
- 2016년 ~ : 한림대학교 국제학부 조교수
- joshua@CybercrimeTech.com

<주관심 분야: 디지털 포렌식, IoT 보안, 빅데이터 분석>

**장 윤 식(정회원)**

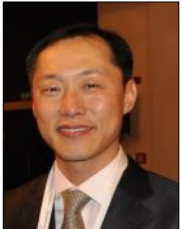

- 2014년 8월 : 고려대 정보경영공학 전문대학원 졸업 (공학박사)
- 2015년 ~ : 한림대학교 국제학부 조교수
- 2005년 2월 ~ 2014년 2월 : 경찰대학 경찰학과 교수
- jakejang@hallym.ac.kr

<주관심 분야: 사이버범죄, 디지털 포렌식, IoT 보안, 범죄분석>

※ This research was supported by Hallym University Research Fund, 2016(HRF-201603-007).